# Family of CV states of definite parity and their metrological power

Mikhail S. Podoshvedov[1,2] and Sergey A. Podoshvedov[1*]

[1]*Laboratory of Quantum Information Processing and Quantum Computing, Institute of Natural and Exact Sciences, South Ural State University (SUSU), Lenin Av. 76, Chelyabinsk, Russia*
[2]*Institute of Physics, Kazan Federal University (KFU), 16a Kremlyovskaya St., Kazan, Russia*
[*]sapodo68@gmail.com

**Abstract:** We introduce a family of new continuous variable (CV) states of definite parity originating from even single-mode squeezed vacuum (SMSV) state by subtracting an arbitrary number of photons from it. A beam splitter (BS) with arbitrary transmittance and reflectance parameters serves as a hub for redirecting input photons in an indistinguishable manner to the output and measuring modes followed by probabilistic measurement, thereby converting the initial SMSV photon distribution into new one after we know the number of registered photons in auxilliary mode. Depending on the parity of the subtracted photons, the family of the generated states is divided into two subfamilies, that is, into even and odd according to the parity of the Fock state subtracted. The family of the CV states is determined solely by one SMSV parameter, which inevitably decreases during their implementation. The algorithm of the quantum engineering can generate macroscopic CV states with a larger average number of photons (say, 4000 and even more) and quantum Fisher information of observable measuring the number of photons many times more (say, from tens of times for small practical values of the squeezing ratio to ≈ 4 for extremely high squeezing amplitude) than one of the original SMSV. The potential of the new family of the CV states of a certain parity, to which original SMSV, no doubt, belongs, can become decisive for a new push to implementation of optical quantum metrology protocols.

## 1. Introduction

Measurement process is a cornerstone for obtaining information from both classical and quantum physical systems realized by assigning a measured value to a physical quantity, thus, giving its estimation. Such an estimate cannot be accurate since some uncertainty is attributed to measured quantity. The error in the measurement result is fundamental (although we note that some statistical error can also be imposed by technical problems) that leaves no chance to describe a quantum system by a single point on the phase plane due to well-known Heisenberg uncertainty relation $\delta x \delta p \geq 0.5$, where $\delta x$ and $\delta p$ are the uncertainties of the position and momentum, respectively, of a quantum particle [1,2]. Although, quantum mechanics imposes fundamental bounds on ultimate achievable precision but at the same time it offers quantum resource (quantum states) to be employed in order to overcome the precision limits, that can be, in principle, obtained using only classical resources [3-12]. Different physical systems can be used to make use of quantum resource for performance of metrological tasks [11,12]. Nevertheless, the photons are the most attractive physical systems for metrological scenarios due to their high mobility and low interaction with environment. It is worth taking into account achievability and practicality of photonic technologies, which includes their generation, manipulation and detection [13-15].

    The goal of photonic quantum metrology is to achieve greater sensitivities (or lower uncertainty) in the measurements of phase-shifts by going beyond classical restriction [16-21] known as standard quantum limit (SQL) or shot-noise (SN). The SQL is determined through

inverse root of average total number of particles. If quantum probe is allowed, the SQL can be overpassed and the phase uncertainty of estimated unknown phase can be reduced down to the quantum boundary known as Heisenberg limit which can improve the precision of classical measurements by a square root of the average number of particles [8-12]. Therefore, it is no occasional that the groundbreaking prediction of quantum mechanics has brought to the fore quantum engineering of high-intensity nonclassical states with the maximum possible Fisher quantum information. In general, we can make use of one of two tricks of quantum mechanics, namely, squeezing [22-24] and entanglement [25-27] to enhance sensitivity above the NS limit. To do it, we develop a quantum engineering algorithm of a new family of nonclassical optical states of definite parity that may have great potential for photonic quantum metrology problems, depending on the chosen observables. The approach includes the BS with arbitrary transmittance and reflectance coefficients capable to redistribute photons of original SMSV to output and measurement modes with the loss of information about from which input Fock state part of the photons is redirected to the measurement mode. Regardless of the measurement outcome in the measuring (auxiliary) mode, a CV state of a certain parity is generated, either even or odd, depending on the parity of the measured Fock state. A whole family of the CV states of definite parity depends solely on one parameter, which largely determines their quantitative nonclassical properties. Although the parameter of the family decreases by the transmittance coefficient squared in the production process compared to the multiplier of the original SMSV, the CV states with an average number of photons in the thousands are observed. Since the mean photon number reflects the energy content of light source, it may indicate the generation of the macroscopic CV states. In addition to their macroscopicity, the CV states can have an uncertainty of the observable measuring the number of photons in a system many times greater than the same uncertainty in the original SMSV which allows for them to decrease quantum Cramer-Rao (QCR) bound of estimation of unknown parameter on compared with QCR bound of the SMSV state, i.e. sub-shot uncertainty of unknown parameter less than one of the original SMSV. The case is not referred to the Heisenberg limit even though sub-Heisenberg sensitivity is observed for the SMSV state. However, the strategy is truly useful for the rapid growth of the variance of the considered observable with the CV states.

## 2. Family of $2m/2m+1$ heralded states and their statistical characteristics

Consider optical scheme in Figure 1 designed to generate CV states of a certain parity. It consists of a lossless beam splitter $BS = \begin{bmatrix} t & -r \\ r & t \end{bmatrix}$, with real transmittance $t$ and reflectance $r$ coefficients satisfying the physical condition $t^2 + r^2 = 1$. The beam splitter is considered to be no longer necessarily balanced and its parameter can be arbitrary [28,29]. A single-mode squeezed vacuum state

$$|SMSV\rangle = \frac{1}{\sqrt{coshs}} \sum_{n=0}^{\infty} \frac{y_0^n}{\sqrt{(2n)!}} \frac{(2n)!}{n!} |2n\rangle \qquad (1)$$

occupies a first mode of the BS, where $y_0 = tanhs/2$ and $s > 0$ is the squeezing parameter of the SMSV that provides the following range of its change $0 \leq y_0 \leq 0.5$. We are going to call $y_0$ SMSV parameter. Zero value of the SMSV parameter $y_0 = 0$ indicates the absence of the SMSV state at the input of the BS, while the value $y_0 = 0.5$ corresponds to the physically unrealizable case of maximally squeezed light with $s \to \infty$. The second mode remains void that is, in a vacuum state $|0\rangle$.

The SMSV passes through the BS $\left(BS_{12}(|SMSV\rangle_1|0\rangle_2)\right)$ with reflection of part of photons into the second auxiliary mode followed by photon number resolving measurement of $n$ photons. Near-unity efficient photon number resolving (PNR) detector is now and experimentally available resource [30,31]. Depending on the parity of the number of detected photons, the measurement outcomes are divided into even $n = 2m$ and odd $n = 2m + 1$. We initially consider an ideal PNR detector with unit quantum efficiency $\eta = 1$. It follows from Eq. (A1), the following $2m$ −heralded states

$$|\Psi_{2m}\rangle = \frac{1}{\sqrt{Z^{(2m)}}}\sum_{n=0}^{\infty} \frac{y_1^n}{\sqrt{(2n)!}} \frac{(2(n+m))!}{(n+m)!} |2n\rangle, \quad (2)$$

are generated provided that even number $2m$ of photons is detected in auxiliary second mode of the BS, where $Z^{(2m)} = d^{2m}Z/dy_1^{2m}$, with $Z \equiv Z(y_1) = 1/\sqrt{1 - 4y_1^2}$ and $y_1 = t^2 \tanh s/2 = t^2 y_0$. The normalization factor is derived in Appendix B (Eqs. (B1-B6)). The parameter $y_1$ changes in the range of $0 \leq y_1 \leq 0.5$, as we adhere to condition $s > 0$. Assume that the PNR detector in the measuring mode registers an odd $2m + 1$ number of photons. Then, the $2m + 1$ heralded state is generated

$$|\Psi_{2m+1}\rangle = \sqrt{\frac{y_1}{Z^{(2m+1)}}}\sum_{n=0}^{\infty} \frac{y_1^n}{\sqrt{(2n+1)!}} \frac{(2(n+m+1))!}{(n+m+1)!} |2n + 1\rangle, \quad (3)$$

where $2m + 1$ odd derivative of the function $Z(y_1)$ on $y_1$ is used in the normalization factor.

In general, the states in Eqs. (2,3) are similar in its form to the original SMSV from which they originate. The difference in the states in Eqs. (1-3) lies in the fact that factor $(2n)!/n!$ of the original SMSV is converted to other multipliers either $\big((2(n + m))!/(n + m)!\big)$ for even or $\big((2(n + m + 1))!/(n + m + 1)!\big)$ for odd CV states involving number $m$ proportional to the number of extracted photons. Additionally, the parameters $y_0$ and $y_1$ differ from each other by a factor $t^2 = y_1/y_0 < 1$, that means that the action BS also leads to a decrease in the input parameter $y_0$. This similarity of the states (1-3) in their form allows us to combine the generated states into one family (class) of the CV states of definite parity. Undoubtedly, even SMSV state from which they come must also belong to the family of the CV states of definite parity. Note a family of the CV states of definite parity and can be formally divided into two groups $2m/2m + 1$ heralded states. This is one of the possible names we are going to use when referring to the generated family. Since the CV states in Eqs. (2,3) consist exclusively of even(odd) Fock states, the $2m$ −heralded state has a certain even parity, while the $2m + 1$ −heralded state is odd CV state. Therefore, the family of the generated states can also be divided on the basis of parity observable into two types or subfamilies, even and odd and, similarly, can also be called even/odd family of CV states.

The condition $y_1 = 0$ formally takes place in the case of either $s = 0$ or $t = 0$ which is not a case for our consideration. It is interesting to note the $2m/2m + 1$ −heralded states become vacuum state $|\Psi_{2m}(y_1 = 0)\rangle = |0\rangle$ and single photon $|\Psi_{2m+1}(y_1 = 0)\rangle = |1\rangle$, respectively, in the case of $y_1 = 0$, regardless of the number of photons $n = 2m, 2m + 1$ subtracted from original SMSV. The opposite case $y_1 = 0.5$ takes place only in the limiting case $t = 1$ and $s \to \infty$ that going beyond the scope of physical consideration. In experimental setup with a high-refractive beam splitter $t \ll 1$ in combination with a small value of the squeezing parameter $s < 1$, the parameter $y_1$ can take on rather small values $y \ll 0.5$. In the case of the minimum value of the measurement parameter $m = 0$, one deals with the CV states of definite parity of the lowest order, namely, the $0 −$ heralded even CV state

$$|\Psi_0\rangle = \frac{1}{\sqrt{Z^{(0)}}}\sum_{n=0}^{\infty} \frac{y_1^n}{\sqrt{(2n)!}} \frac{(2n)!}{n!} |2n\rangle, \quad (4)$$

and $1 −$ heralded odd CV state

$$|\Psi_1\rangle = \sqrt{\frac{y_1}{Z^{(1)}}}\sum_{n=0}^{\infty} \frac{y_1^n}{\sqrt{(2n+1)!}} \frac{(2(n+1))!}{(n+1)!} |2n + 1\rangle. \quad (5)$$

It is interesting to note the state (1) may recall in its form the original SMSV state. The only action of the BS with no click at auxiliary mode is to change SMSV parameter $y_0$ onto $y_1$ ($y_0 \to y_1$) to implement the CV state in Eq. (4). When $y_1 = y_0$, the state (4) definitely becomes SMSV in Eq. (1), i.e. $|\Psi_0\rangle \equiv |SMSV\rangle$.

Extraction of a certain number of photons either $2m$ or $2m+1$ from the SMSV in an indistinguishable manner with loss of all information from which Fock states of the original superposition the photons are subtracted generates $2m/2m+1-$ heralded states (2-5) leaving the output state with a well-defined parity either even or odd in dependency on the parity of the measurement outcome. The success probabilities to implement the $2m-$(even) and $2m+1-$(odd) heralded CV states

$$P_{2m} = \frac{1}{\cosh s}\left(\frac{1-t^2}{t^2}\right)^{2m}\frac{y^{2m}}{(2m)!}Z^{(2m)}, \quad (6)$$

$$P_{2m+1} = \frac{1}{\cosh s}\left(\frac{1-t^2}{t^2}\right)^{2m+1}\frac{y^{2m+1}}{(2m+1)!}Z^{(2m+1)}, \quad (7)$$

respectively, are subject to the normalization condition $\sum_{m=0}^{\infty}(P_{2m}+P_{2m+1}) = 1$ that is directly verified. Note that the states in Eqs. (1-5) depend solely on one parameter either $y_1$ or $y_0$, while the success probabilities are determined by two parameters, namely, $y_1(y_0)$ and BS parameter $(1-t^2)/t^2$. The parameter $(1-t^2)/t^2$ raised to some power can only give substantially decreasing multiplier into the success probability in the case of $t > 1/\sqrt{2}$ and especially in the case of use of highly transmitting BS ($t \to 1$). Nevertheless, note that the BS factor can be increasing multiplier for the success probability if $t < 1/\sqrt{2}$. Dependences of the success probabilities on the squeezing parameter $s$ for different values of the BS transmittance coefficient $t$ are shown in Figure 2. Note that the success probability of the state (4) clearly prevails over other probabilities, especially with $t$ growing i.e. when $t \to 1$. This is due to the fact that a significant part of the photons passes through the BS in the case of a highly transmitting BS with $t \to 1$, and then it is more likely that no click will be registered in auxiliary mode. The success probabilities of the $2m/2m+1-$ heralded CV states with $m \neq 0$ take smaller values and their peaks are shifted towards larger values of the squeezing amplitude $s$.

Knowing the exact form of the $2m/2m+1-$ heralded states, one can find all their statistical characteristics: average number of photons $\langle n \rangle$, the average number of photons squared $\langle n^2 \rangle$ and variation $\delta n$. Using equations (2,3), one obtains

$$\langle n \rangle_{2m} = y\frac{Z^{(2m+1)}}{Z^{(2m)}}, \quad (8)$$

$$\langle n^2 \rangle_{2m} = y^2\frac{Z^{(2m+2)}}{Z^{(2m)}} + \langle n \rangle_{2m}, \quad (9)$$

$$\delta n_{2m} = \langle n^2 \rangle_{2m} - \langle n \rangle_{2m}^2 = y^2\frac{Z^{(2m+2)}}{Z^{(2m)}} + \langle n \rangle_{2m} - \langle n \rangle_{2m}^2, \quad (10)$$

for even CV states and

$$\langle n \rangle_{2m+1} = y\frac{Z^{(2m+2)}}{Z^{(2m+1)}}, \quad (11)$$

$$\langle n^2 \rangle_{2m+1} = y^2\frac{Z^{(2m+3)}}{Z^{(2m+1)}} + \langle n \rangle_{2m+1}, \quad (12)$$

$$\delta n_{2m+1} = \langle n^2 \rangle_{2m+1} - \langle n \rangle_{2m+1}^2 = y^2\frac{Z^{(2m+3)}}{Z^{(2m+1)}} + \langle n \rangle_{2m+1} - \langle n \rangle_{2m+1}^2. \quad (13)$$

for odd CV ones. In general case, we define the statistical parameters with parameter $y$. For the states Eqs. (2,3), $y$ should be replaced by $y_1$ and the derivative of the function $Z(y_1)$ is also taken over the $y_1$. The statistical characteristics are mainly determined by the respective derivatives of the function $Z(y_1)$, which allows them to be evaluated in clearer form. The analytical forms of the statistical characteristics when we calculate the higher order derivatives with $m > 0$ become rather complicated in terms of $y_1$ and $Z(y_1)$. Let us only analytically derive expressions for average number of photons $\langle n \rangle_0, \langle n \rangle_1$ and variations $\delta n_0, \delta n_1$ corresponding to minimum value of $m = 0$. So, we have $\langle n \rangle_0 = 4y_1^2 Z_1^2(y_1)$, $\delta n_0 = 2(\langle n \rangle_0^2 + \langle n \rangle_0)$ for the state in Eq. (4) and $\langle n \rangle_1 = 1 + 12y_1^2 Z_1^2(y_1)$, $\delta n_1 = 2(\langle n \rangle_1 - 1)(\langle n \rangle_1 + 2)/3 = 2(\langle n \rangle_1^2 + \langle n \rangle_1 - 2)/3$ for the state in Eq. (5). It follows from the expressions that $\langle n \rangle_0(y_1 = 0) = 0$ and $\delta n_0(y_1 = 0) = 0$ but $\langle n \rangle_1(y_1 = 0) = 1$ and $\delta n_1(y_1 = 0) = 4/3$.

So, using the expressions for $\langle n \rangle_0$ and $\delta n_0$, one can estimate average number of photons and variance of the SMSV by taking $t = 1$ in $y_1$ that leads to $y_1 = y_0$ and gives $\langle n \rangle_{SMSV} = sinh^2 s$ and $\delta n_{SMSV} = 2(\langle n \rangle_0^2 + \langle n \rangle_0) = 2(sinh^4 s + sinh^2 s)$. The average number of photons in SMSV state is proportional to the function $sinh s$ and can grow exponentially with increasing the squeezing parameter $s$. So, we have $\langle n \rangle_{SMSV} = 0.2704$ for $s = 0.5$, $\langle n \rangle_{SMSV} = 1.3924$ for $s = 1$, $\langle n \rangle_{SMSV} = 4.5369$ for $s = 1.5$, $\langle n \rangle_{SMSV} = 13.1769$ for $s = 2$, $\langle n \rangle_{SMSV} = 100.4$ for $s = 3$, $\langle n \rangle_{SMSV} = 744.74$ for $s = 4$ and $\langle n \rangle_{SMSV} = 5505.64$ for $s = 5$. Although at present none of the measures is recognized as a generally accepted measure of macroscopicity, in general, the SMSV state with average number of photons $> 1000$ may be recognized as macroscopic and may have a serious potential in precise measurement. But increase of average number of photons $\langle n \rangle_{SMSV}$ requires a significant increase of the squeezing parameter $s$, which can hardly be achieved in practice. As a rule, in experimental setups, the SMSV cannot be considered to be macroscopic, since the peak of its distribution is shifted towards smaller Fock states.

Let us compare statistical characteristics even/odd CV states in Eqs. (2,3) with ones of SMSV. As a comparison, we choose the average number of photons $\langle n \rangle_n$, the ratio of the average number of photons in $2m/2m + 1 -$ heralded and SMSV states i.e. $Rn_n = \langle n \rangle_n / \langle n \rangle_{SMSV} = \langle n \rangle_n / sinh^2 s$, square root of the variance or the same photon uncertainty $\sqrt{\delta n_n}$ and the ratio of the photon uncertainty of the generated CV state of definite parity to one of the SMSV, i.e. $RV_n = \sqrt{\delta n_n}/\sqrt{\delta n_{SMSV}} = \sqrt{\delta n_n}/\sqrt{2(sinh^4 s + sinh^2 s)}$, where the subscript $n$ takes on either even $n = 2m$ or odd $n = 2m + 1$ values. In Figures 3, 4, we show the dependencies of the $\langle n \rangle_n$, $Rn_n$, $\sqrt{\delta n_n}$ and $RV_n$ on the squeezing parameter $s$ for two parameters $t = 0.98$ (Fig. 3) and $t = 0.99$ (Fig. 4). The figures 3(a), 4(a) show the average number of photons $\langle n \rangle_{SMSV}$ in SMSV state in black. Also figures 3(c), 4(c) display uncertainty $\sqrt{\delta n_{SMSV}}$ in black for comparison with other $\sqrt{\delta n_n}$.

As can be seen from the plots in the figures 3(a,b), 4(a,b), the average number of photons $\langle n \rangle_n$ of the $2m/2m + 1 -$ heralded states of definite parity exceeds the number of photons $\langle n \rangle_{SMSV}$ i.e. $\langle n \rangle_n > \langle n \rangle_{SMSV}$ for a large number of $n$ in a wide range of variation of the squeezing parameter $s$, especially, for the case of $t = 0.99$. Although it is worth noting that there are cases when $\langle n \rangle_{SMSV} > \langle n \rangle_n$ for some values of $n$ and $s$. A strong predominance $\langle n \rangle_n \gg \langle n \rangle_{SMSV}$ is observed for the case of $s \ll 1$ that is not of interest, since, on the whole, the average number of photons $\langle n \rangle_n$ is small in the range. Maximally observed value of the average number of photons is $\langle n \rangle_{100} \approx 4000$ in the case of the subtraction of 100 photons from original SMSV with squeezing amplitude of $s = 3$ in the case of $t = 0.99$ (Fig. 4(a)). It can indicate evident macroscopic properties of the $100 -$heralded state on compared with original SMSV, i.e. a non-Gaussian procedure of subtraction of a large number of photons, say 100, can redistribute the remaining photons so that the average number of photons in the generated state can exceed the average number of photons in the initial SMSV by 40 times $\langle n \rangle_{100} \approx 4000 > \langle n \rangle_{SMSV} = 100.4$. The parameter $Rn_n$ behaves accordingly. With an increase in the squeezing parameter $s$, the ratio $Rn_n$ decreases at any value of $n$ and $t$, nevertheless, remaining more than one, i.e. $Rn_n > 1$, except for small values of $n$ (Figs. 3(b), 4(b)), which indicates that the number of average photons in $2m/2m + 1 -$generated states keeps greater than in the initial SMSV from which they originate, especially, when high-transmission beam splitter is used. Note that the ratio can be even higher than 10, i.e. $Rn_n > 10$, in the case of $s = 3$ (Fig. 4(b)). As for the photon uncertainty of the $2m/2m + 1 -$ heralded CV states, as can be seen from figures 3(c) and 4(c), the parameter $\sqrt{\delta n_n}$ takes on a larger values in comparison with one of original SMSV shown by black line in Figs. 3(c) and 4(c)), i.e. $\sqrt{\delta n_n} > \sqrt{\delta n_{SMSV}}$ in the vast majority of cases considered. Note that the use of the BS with $t = 0.99$ (Fig. 4(c)) allows for one to obtain significantly larger values of $\sqrt{\delta n_n}$ on compared with the case of $t = 0.98$ (Fig. 3(c)). The maximum value of uncertainty $\sqrt{\delta n_n} \approx 400$ observed in Figure 4(c) corresponds to the state $|\Psi_{100}\rangle$ at $s = 3$. An inequality $\sqrt{\delta n_n} > \sqrt{\delta n_{SMSV}}$ is observed in the overwhelming majority of cases that provides $RV_n > 1$ for them. So, the ratio $RV_{100}$ can take on values slightly less than 4 in the case of $s = 3$ (Fig. 4(d)), that, together with substantial increase in photon uncertainty $\sqrt{\delta n_n}$, can have potential for application in quantum metrology exceeding possibilities of the original SMSV state in estimation of unknown parameter with sub-shot-noise uncertainty. The ratio can only increase due to a decrease in the squeezing amplitude $s$. For example, $RV_{100}$ becomes close to 10 for $s = 2$ in Fig. 4(d)), which, nevertheless, reduces absolute value of $\sqrt{\delta n_n}$.

The family of the $2m/2m + 1 -$ heralded states stems from nonclassical SMSV state which possesses pronounced nonclassical properties, namely, the noise in one of the two quadrature components $X_1 = (a + a^+)/2$ and $X_2 = (a - a^+)/2i$, where $a^+, a$ are the creation and annihilation operator, can be less than the noise of the vacuum state. Using exact form of the CV states in Eqs. (2,3), one can directly derive analytic compact expressions of the quadrature variances $(\Delta^2 X_1)_n = \langle X_1^2 \rangle_n - \langle X_1 \rangle_n^2$ and $(\Delta^2 X_2)_n = \langle X_2^2 \rangle_n - \langle X_2 \rangle_n^2$

$$(\Delta^2 X_1)_{2m,2m+1} = \frac{1}{4} + \frac{\langle n \rangle_{2m,2m+1}}{2} + \frac{y_1}{Z^{(2m,2m+1)}} (y_1 Z)^{(2m+1,2m+2)}, \quad (14)$$

$$(\Delta^2 X_2)_{2m,2m+1} = \frac{1}{4} + \frac{\langle n \rangle_{2m,2m+1}}{2} - \frac{y_1}{Z^{(2m,2m+1)}} (y_1 Z)^{(2m+1,2m+2)}, \quad (15)$$

where the number $n$ is either even $n = 2m$ or odd $n = 2m + 1$. Note that the form is more convenient than if we represented the derivatives in terms of $y_1$ and $Z(y_1)$. As can be seen from the expressions, the difference between the two quadrature variances is only in the last term, which is present either with sign + or with − sign. Obviously, quadrature variances $(\Delta^2 X_2)_n$ take a smaller value compared to $(\Delta^2 X_1)_n$, where $n$ can be either even $n = 2m$ or odd $n = 2m + 1$ provided that the derivatives $(y_1 Z)^{(2m+1,2m+2)}$ take positive values. Note that the quadrature variances for the SMSV directly follow from Eqs. (14,15), i.e. $(\Delta^2 X_1)_{SMSV} = exp(2s)/4 \geq 1/4$ (SMSV desqueezing) and $(\Delta^2 X_2)_{SMSV} = exp(-2s)/4 \leq 1/4$ (SMSV squeezing) if we take $m = 0$ and $y_1 = y_0$ in Eqs. (14,15). Product of two quadrature uncertainties is $(\delta X_1)_{SMSV}(\delta X_2)_{SMSV} = 1/4$, where $(\delta X_1)_{SMSV} = \sqrt{(\Delta^2 X_1)_{SMSV}}$ and $(\delta X_2)_{SMSV} = \sqrt{(\Delta^2 X_2)_{SMSV}}$. Thus, the presence of the last term with sign − in Eq. (15) may lead to the observation of the quadrature squeezing, i.e. noise reduction in comparison with vacuum one $(\delta X)_n \equiv (\delta X_2)_n = \sqrt{(\Delta^2 X_2)_n} < 0.5$, where the subscript $n$ may take on either even $n = 2m$ or odd $n = 2m + 1$ integers. So, one can directly output the quadrature variances in simple terms $(\Delta^2 X_2)_0 = 1/4 - y_1/(1 + 2y_1)$ and $(\Delta^2 X_1)_0 = 1/4 + y_1/(1 + 2y_1)$ for even $0 -$heralded state (4) from Eqs. (14,15) and only quadrature uncertainty $(\delta X_2)_0$ can only demonstrate quadrature squeezing. Consider the behavior of the quadrature components in the case of $y_1 = 0$. The quadrature uncertainty for even CV states takes on the value $(\delta X)_{2m}(y_1 = 0) = 0.5$ as $\langle n \rangle_{2m}(y_1 = 0) = 0$. In the case of odd CV states taken from Eq. (3), the quadrature uncertainty changes $(\delta X)_{2m+1}(y_1 = 0) = \sqrt{0.75}$ $((\Delta^2 X_2)_{2m+1} = 0.75)$ as $\langle n \rangle_{2m+1}(y_1 = 0) = 1$ which is related to the aforementioned approximation of $2m + 1 -$heralded states by a single photon in the case of $y_1 = 0$. Thus, it is worth considering separately quadrature squeezing for $2m -$ and $2m + 1 -$heralded states.

We show in figures (5,6) dependencies of the quadrature squeezing $(\delta X)_{2m}$ for $2m -$ (Figs. 5(a),6(a)) and $(\delta X)_{2m+1}$ for $2m + 1 -$ (Figs. 5(b),6(b)) heralded states as well as quadrature squeezing $(\delta X)_{SMSV}$ (blue line) for the SMSV itself on the squeezing amplitude $s$ of the original SMSV. The plots in figures 5 and 6 are constructed for the BS parameters $t = 0.9$ and $t =$ 0.99, respectively. The plots in figs. 5(c) and 6(c) show dependencies of $RS_{2m}$ being ratio of the quadrature squeezing of the CV states of definite parity to SMSV quadrature squeezing $RS_{2m} = (\delta X)_{2m}/(\delta X)_{SMSV}$ on the squeezing amplitude $s$. Ratios of $RS_{2m+1} = (\delta X)_{2m+1}/(\delta X)_{SMSV}$ in dependency on $s$ are depicted in figs. 5(d) and 6(d). As can be seen from the figures, the generated states can exhibit the quadrature squeezing $(\delta X)_n < 0.5$. But it is worth noting that the quadrature squeezing of $2m-$ heralded states (Figs. 5(a),6(a)) is observed $((\delta X)_{2m} < 0.5)$ for any value of $m$ in the entire range of the squeezing amplitude $s$. The quadrature squeezing $(\delta X)_{2m}$ experiences both local minimum and local maximum at certain values of $s < 1$, that can entail generating more squeezed states as compared to the initial SMSV, i.e. $(\delta X)_{2m} < (\delta X)_{SMSV}$ or the same $RS_{2m} < 1$ for a given range of variation of $s$. While for odd CV states, there are values of $s < 1$ at which the quadrature component $(\delta X)_{2m+1}$ takes on values greater than 0.5 i.e. $(\delta X)_{2m+1} > 0.5$. One should also note that for all values of the squeezing amplitude $s$, the condition $(\delta X)_{2m+1} > (\delta X)_{SMSV}$ is satisfied, which guarantees $RS_{2m+1} > 1$.

## 3. Phase estimation with $2m/2m+1$ heralded states

Engineering of macroscopic nonclassical states that can have large quantum Fisher information for a particular observable is the cornerstone for ultra-precise estimation of an unknown parameter [8-10,18-21], in our case, unknown phase $\varphi$. In general, quantum Fisher information is difficult to compute under general scenario with arbitrary quantum channel $\Phi_\varphi(\rho) = \rho(\varphi)$ that transforms an initial state $\rho$ into encoded with an unknown parameter $\rho(\varphi)$ already carrying information about an unknown parameter. Nevertheless, the problem is greatly simplified by considering only unitary encoding $\rho(\varphi) = U_\varphi \rho U_\varphi^+$, where $U_\varphi = exp(-i\varphi G)$ is a unitary operator providing the parameter encoding, $U_\varphi^+$ is Hermitian conjugate and $G$ is a Hermitian operator often named the generator of the unitary transformation. Let us consider a very simple choice of the generator $G = n/2$, where the number operator $n = a^+ a$ is Hermitian observable measuring the number of photons in a system (in a state). Corresponding unitary encoding is given by $U_\varphi = exp(-i\varphi n/2)$ which causes clockwise rotation by angle $\varphi/2$ on phase plane that induces change in the phase of input state $\rho(\varphi)$ relative to the initial state $\rho$.

In the case of the pure states in equations (2,3), one can calculate QCR bound of estimation of unknown parameter $\varphi$ in terms of the variance of the corresponding observable [18-21]. Indeed, the Fisher quantum information is related to the dispersion of the generator as $F(|\Psi_n\rangle, G) = 4\Delta^2 G$ which in the case considered gives

$$F(|\Psi_n\rangle, n/2) = \delta n_n = y^2 \frac{Z^{(n+2)}}{Z^{(n)}} + y \frac{Z^{(n+1)}}{Z^{(n)}} - \left(y \frac{Z^{(n+1)}}{Z^{(n)}}\right)^2, \qquad (16)$$

where the subscript $n$ can take both even $n = 2m$ and odd $n = 2m+1$ integers. It is worth heeding that $y$ must be replaced by $y_1$ ($y \to y_1$) in case of using the states in Eqs. (2,3). It allows us to calculate the QCR bound [18-21] for estimation of unknown parameter $\varphi$ as $\Delta\varphi_{QCR,n} = 1/\sqrt{F(|\Psi_n\rangle, G = n/2)}$ and $\Delta\varphi_{QCR,SMSV} = 1/\sqrt{F(|SMSV\rangle, G = n/2)}$, that is, the minimal estimation error. Obviously, the smaller the estimation error, the more accurate the procedure for estimating the unknown parameter can be. The graphs in the figures (3,4) can be adapted to estimate uncertainty bound of the unknown parameter $\varphi$ as the generated even/odd CV states of definite parity can have statistical characteristics suitable for it. Figures 7(a-d) show the values of the QCR bound $\Delta\varphi_{QCR,n}$, $\Delta\varphi_{QCR,SMSV}$ and $g_n = -10log_{10}(\Delta\varphi_{QCR,n}/\Delta\varphi_{QCR,SMSV})$ being the sensitivity gain measured in $dB$ depending on the squeezing amplitude s of the initial SMSV state. The curves in Figure 7(a,c) are plotted for $t = 0.98$, and the graphs in Figure 7(b,d) are for $t = 0.99$. The curves $\Delta\varphi_{QCR,SMSV}$ in Figures 7(a,b) are shown in black. As can be seen from the dependencies, there is a decrease in $\Delta\varphi_{QCR,n}$ compared to $\Delta\varphi_{QCR,SMSV}$, i.e. $\Delta\varphi_{QCR,n} < \Delta\varphi_{QCR,SMSV}$ for the overwhelming number $n$. The reverse condition $\Delta\varphi_{QCR,n} > \Delta\varphi_{QCR,SMSV}$ is observed for small values of $n$ and large values of the squeezing amplitude $s$. The QCR bound can reach the values of $\Delta\varphi_{QCR,n} \sim 0.1$ in the case of $s \ll 1$ and can decrease by an order of magnitude to $\Delta\varphi_{QCR,n} \sim 0.01$ in the case of the squeezing amplitude $s = 3$. As a result, the sensitivity gain becomes greater than one, that is, $g_n > 1$ is observed in the vast majority of cases. Moreover, the sensitivity gain $g_n$ can take values from $g_n = 20$ for $s \ll 1$ to $g_n \approx 4.8$ in the case of large values of the squeezing amplitude $s = 3$ in the case of $t = 0.99$.

## 4. Conclusion

We have introduced into consideration a new class of the CV states of definite parity arising from original SMSV state. The mechanism of their implementation is based on subtracting a certain number of photons from the SMSV by redirecting them indistinguishably into output and measurement modes by the BS hub and measuring the number of photons in measuring mode. The subtraction of photons from original SMSV in indistinguishable manner redistributes input photon distribution of the SMSV state. The photon redistribution by eliminating the lower-photon states contribution, involving vacuum, which can have a larger weight in the initial SMSV distribution generates a family of CV states of definite parity in Eqs. (2-5). Difference between the forms of the generated states and the original SMSV is rather insignificant. They are all determined by one parameter either $y_1$ in the case of generated states of definite parity or $y_0$ for SMSV which are interconnected to each other. The parameter $y_1$ inevitably decreases by BS transmittance coefficient squared in the process of extraction of any number of photons from the SMSV state. The family of the $2m/2m+1-$ heralded states also involves a factor indicating the number of extracted photons $m$ unlike original SMSV. The factor shifts the initial SMSV photon distribution towards multiphoton states that can lead to an increase in the average number of photons and photon uncertainty. In addition, the generated even/odd CV states can also possess squeezed noise in one of the quadrature components, which can also indicate their similarity with the original SMSV. The consonance between $2m/2m+1-$ heralded states and SMSV state enables to involve it to the family of the CV states of definite parity.

The average number of photons in the generated states can exceed tenfold the average number of photons in the initial SMSV which allows us to speak about the macroscopic nature of the generated CV states. This circumstance can be of practical importance as a method for generating macroscopic (bright) nonclassical CV states of a certain parity. For example, even using the initial SMSV state with a squeezing amplitude $s = 1.5$ with average number of photons $\langle n \rangle_{SMSV} = 13.18$ can already generate nonclassical states with an average number of photons of several hundred. The gain in brightness of the nonclassical CV states is feasible and can find new applications. Additionally, due to the high photon uncertainty of the generated CV states of a certain parity, there is a serious potential for them to be used in more accurate estimation of unknown parameter compared not only with classical states but also with nonclassical ones. We have shown that QCR bound of the CV states of certain parity can take smaller values in comparison with the QCR bound of the SMSV state even despite its sub-Heisenberg sensitivity, when observable measuring the number of photons in the optical mode is chosen. In the case of the CV states of certain parity, we do not talk about reaching the Heisenberg limit, since the average number of photons squared exceeds the dispersion of the number of photons, i.e. $\langle n \rangle_n^2 > \delta n_n > \langle n \rangle_n$ for $n > 0$. Nevertheless, in a considered squeezing amplitude range from 0 to 3, the uncertainty in the number of photons can take on values much larger than the photon uncertainty in the initial SMSV state. In general, the developed approach feasible in practice with a certain parity state, at least with a small number $n = 0 - 10$, can actually be implemented in practice. Note that the method of subtracting photons from different input states in different configurations is quite widely used in scientific research [28,29]. A natural obstacle to a more efficient application of the method may be input SMSV state with insufficiently high squeezing amplitude. In particular, a light with $15dB$ squeezing corresponding to the squeezing amplitude of $s \approx 1.7$ and implemented in practice [32] may even not be so sufficient fully to demonstrate the efficiency of the method. However, the approach can be extended by using a hub with several BSs and the same number of the PNR detectors. Then, the SMSV state with a small squeezing amplitude could be transformed into the CV state of a certain parity, provided that certain measurement outcomes are registered in the measurement modes. It is also interesting to consider the possibilities of the CV states of a certain parity for estimating an unknown parameter in other optical schemes, in particular, in the Mach-Zender interferometer, which is a topic for a separate study.

**Appendix A. Derivation of the states in Eqs. (2-5)**

The analytical derivation of the $2m/2m + 1-$ heralded states in Eqs. (2-5) starts with transformations of the creation operators imposed by the beam splitter: $a_1^+ \to ta_1^+ - ra_2^+, a_2^+ \to ra_1^+ + ta_2^+$ that realize the following unitary transformation over input SMSV due to linearity of the BS operator

$$BS_{12}(|SMSV\rangle_1|0\rangle_2) = \frac{1}{\sqrt{\cosh s}} \sum_{n=0}^{\infty} C_n |\Psi_n\rangle_1 |n\rangle_2, \tag{A1}$$

where the amplitudes $C_n$ are determined as

$$C_n = (-1)^n \left(\frac{1-t^2}{t^2}\right)^{\frac{n}{2}} \frac{y_1^{\frac{n}{2}}}{\sqrt{n!}} \begin{cases} \sqrt{Z^{(2m)}(y_1)}, if \ n = 2m \\ \sqrt{Z^{(2m+1)}(y_1)}, if \ n = 2m+1 \end{cases}. \tag{A2}$$

Measurement of an even number $n = 2m$ of photons in the second auxiliary mode makes it possible to realize even CV states in Eqs. (2,4), while the if the observer registers an odd number $n = 2m + 1$ of photons, then the heralded CV state becomes odd as in Eqs. (3,5). The success probabilities to realize the $2m/2m + 1$ heralded states are the following $P_n = |C_n|^2/\cosh s$ (Eqs. (6,7)) satisfying the normalization condition $\sum_{n=0}^{\infty} |C_n|^2/\cosh s = 1$.

**Appendix B. Formula derivation for $Z(y)$**

Mathematical inference of the function $Z(y)$, which is the basis for the derivation of the normalized factors of the $2m/2m + 1$ heralded states as well as for their statistical characteristics, can start with inner product of the squeezed even state $|SS\rangle = \sum_{n=0}^{\infty} a_{2n} |2n\rangle$ like in Eq. (1) with normalization condition $\sum_{n=0}^{\infty} |a_{2n}|^2 = 1$ written in the $x \ (coordinate)-$ representation $\Psi_{SS}(x) = \langle x|SS\rangle$ and the coherent state $|\alpha\rangle = exp(-|\alpha|^2/2) \sum_{n=0}^{\infty} (\alpha^n/\sqrt{n!})|n\rangle$ with real amplitude $\alpha$ also presented in $x-$ representation $\Psi_\alpha(x) = \langle x|\alpha\rangle$

$$\langle \alpha|SS\rangle = \int_{-\infty}^{\infty} \Psi_\alpha(x) \Psi_{SS}(x) dx = \sqrt{\frac{2R}{1+R^2}} exp\left(-\frac{R^2 \alpha^2}{1+R^2}\right), \tag{B1}$$

where

$$\Psi_{SS}(x) = \frac{\sqrt{R}}{\pi^{1/4}} exp\left(-\frac{R^2 x^2}{2}\right), \tag{B2}$$

with the real parameter $R$ responsible for state's "squeezing" properties and

$$\Psi_\alpha(x) = \frac{1}{\pi^{1/4}} exp\left(-\frac{(x-\sqrt{2}\alpha)^2}{2}\right). \tag{B3}$$

Expanding the right expression in a Taylor series in Eq. (B1), one obtains

$$a_{2n} = \langle 2n|SS\rangle = \sqrt{\frac{2R}{1+R^2}} \left(\frac{1-R^2}{2(1+R^2)}\right)^n \frac{\sqrt{(2n)!}}{n!}, \quad (B4)$$

which can be rewritten in terms of new parameter $y = (1-R^2)/(2(1+R^2))$ as

$$a_{2n} = \frac{y^n \sqrt{(2n)!}}{\sqrt{Z(y)} n!}, \quad (B5)$$

where the normalization factor

$$Z(y) = \frac{1+R^2}{2R} = \frac{1}{\sqrt{1-4y^2}} \quad (B6)$$

is introduced. In particular, if we choose $R = exp(-s)$, then we have $y = tahhs/2$ that gives SMSV amplitudes: $a_{2n} = b_{2n}/\sqrt{coshr}$ (Eq. (1)).

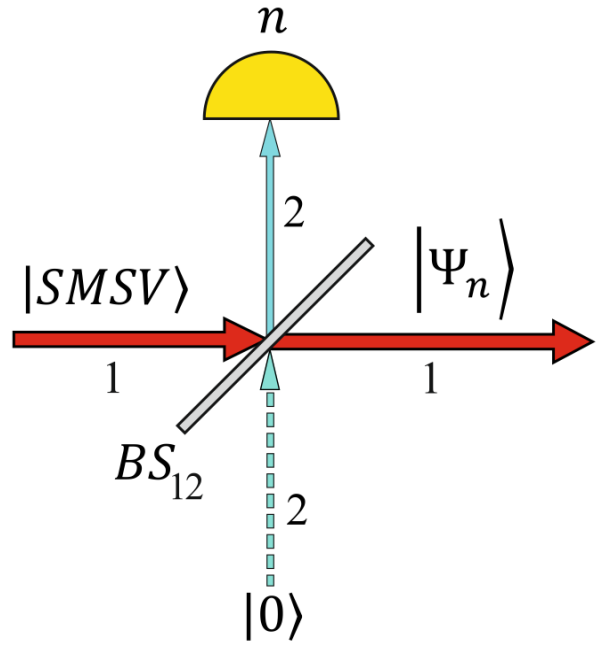

**Figure 1** An optical scheme used to generate CV states of definite parity either even or odd. The hub consists of the beam splitter with arbitrary transmittance coefficient $t$ through which the original SMSV with squeezing amplitude $s$ passes and PNR detector. Second measurement mode is used to measure number of photons and implement new measurement induced CV states $|\Psi_n\rangle$, where either $n = 2m$ or $n = 2m + 1$ is used.

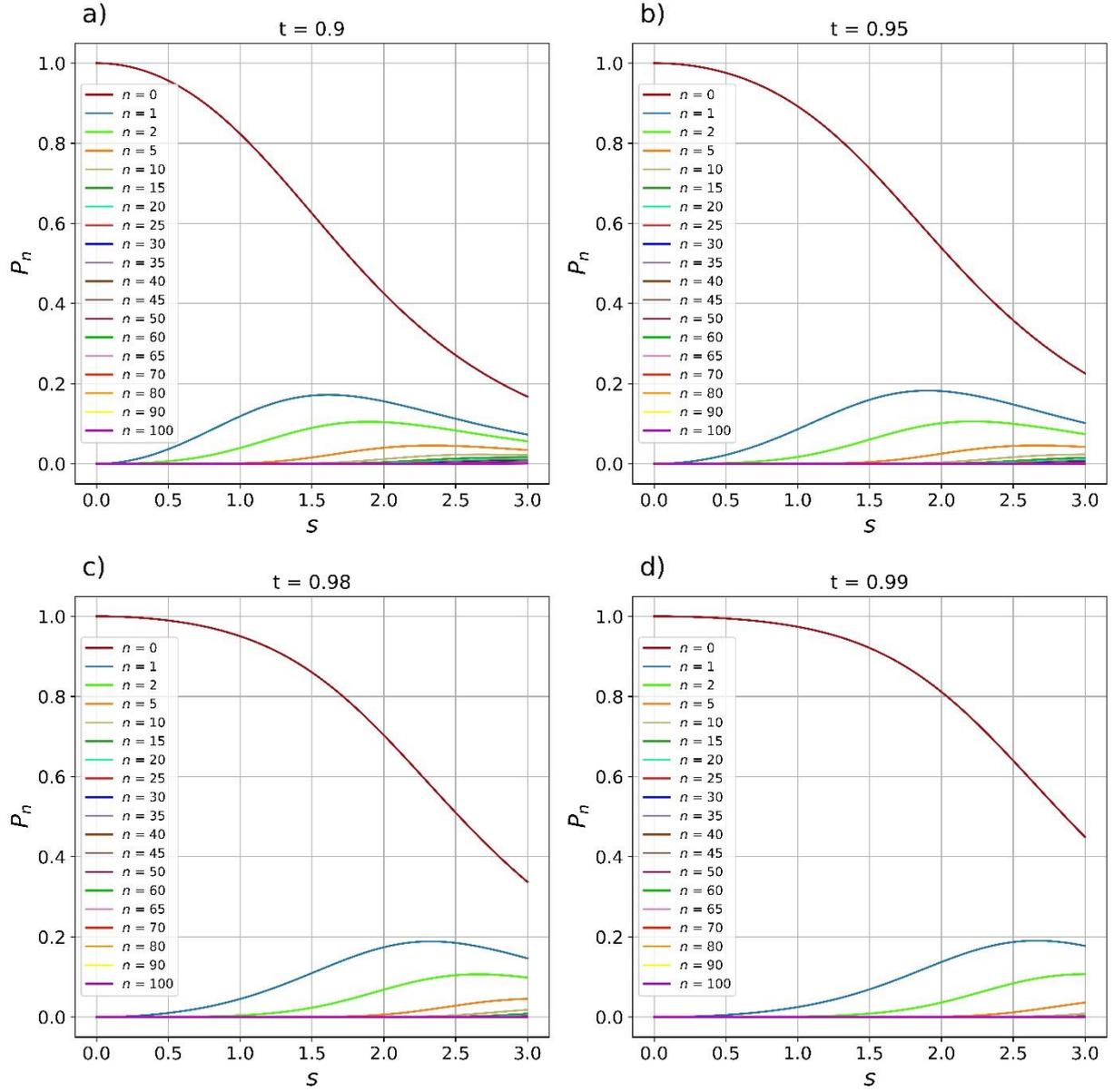

**Figure 2(a-d)** Dependencies of the success probabilities $P_n$ (Eqs. (6,7)) on the squeezing amplitude $s$ of the original SMSV for different values of the BS parameter $t$. The success probability $P_0$ prevails over others ($P_0 > P_n, n > 0$) over the entire range of values of $s$ for different values of $t$. Peak of the success probabilities shift towards larger values of $s$ with an increase in the BS parameter $t \to 1$.

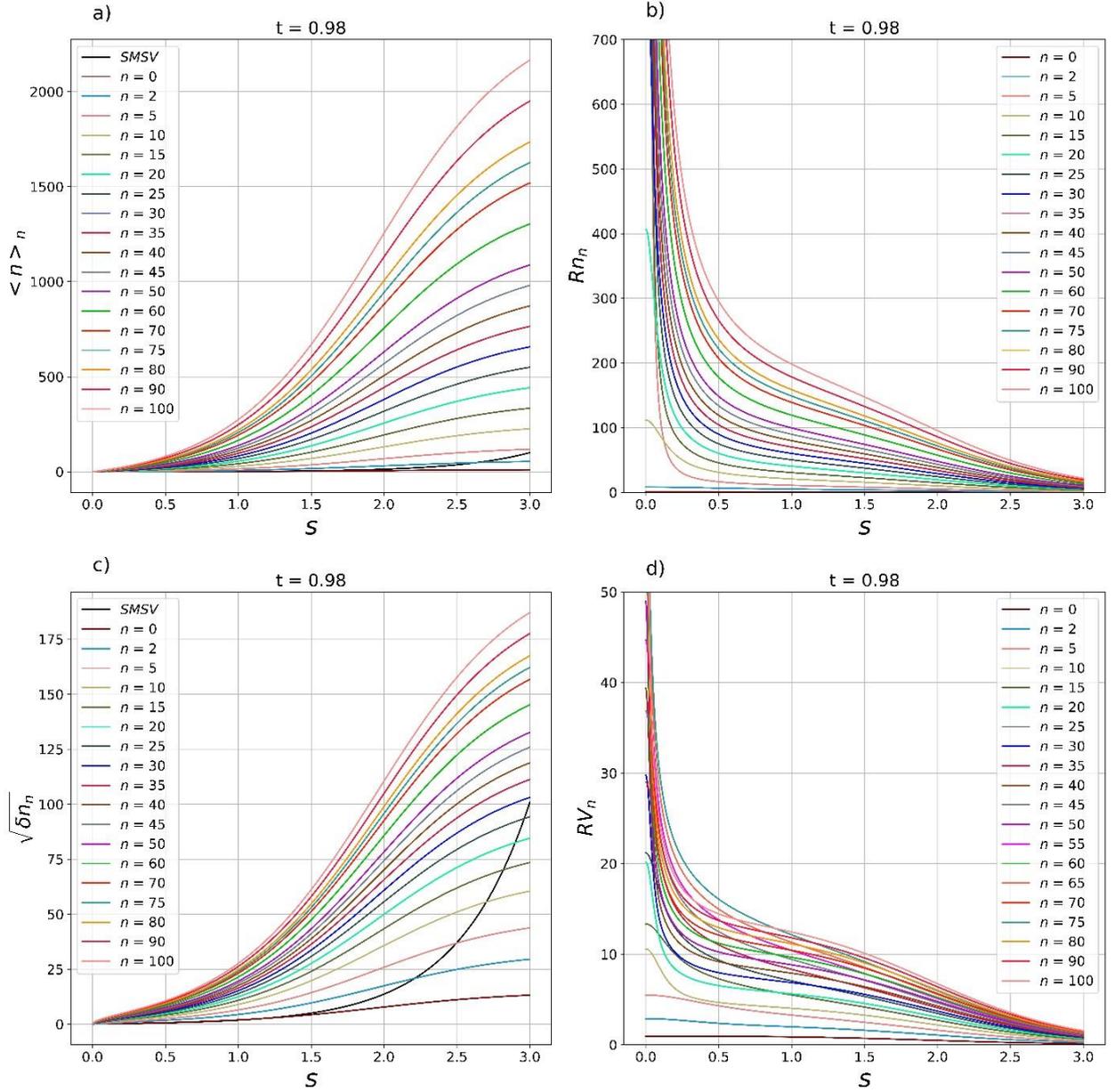

**Figure 3(a-d)** Dependencies of (a) average number of photons $\langle n \rangle_n$, (b) ratio $Rn_n = \langle n \rangle_n / \langle n \rangle_{SMSV} = \langle n \rangle_n / sinh^2 s$ (b), (c) photon uncertainty $\sqrt{\delta n_n}$ and (d) ratio $RV_n = \sqrt{\delta n_n}/\sqrt{\delta n_{SMSV}} = \sqrt{\delta n_n}/\sqrt{2(sinh^4 s + sinh^2 s)}$ on the squeezing amplitude $s$ for the BS parameter $t = 0.98$. Here, integer $n$ is changed from $n = 0$ up to $n = 100$ and in (a) $\langle n \rangle_{SMSV}$ and in (c) $\sqrt{\delta n_{SMSV}}$ dependencies are highlighted in black.

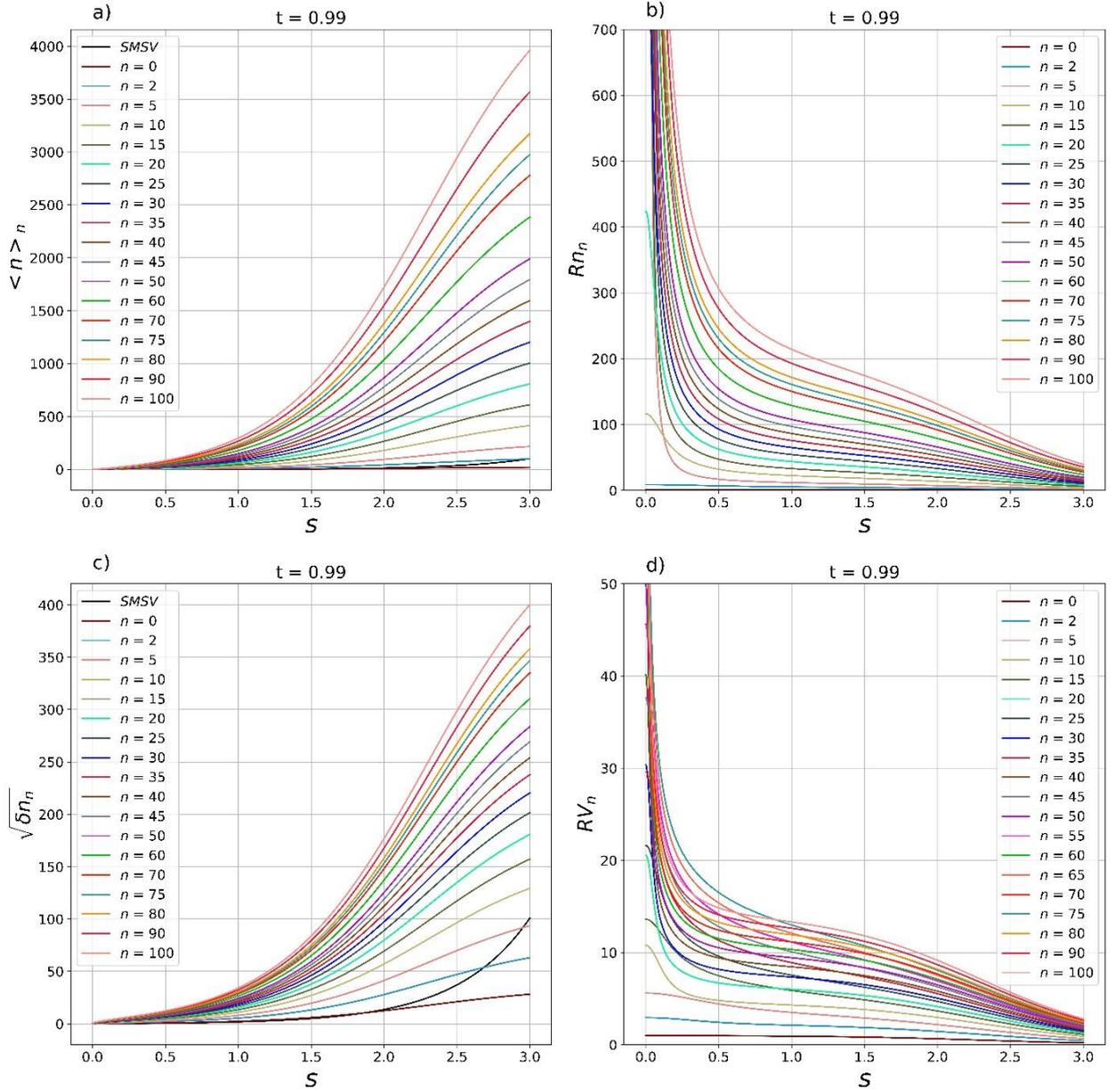

**Figure 4(a-d)** The same dependencies as in the case of Figure 3 but with $t = 0.99$. Comparing the plots in Figures 3 and 4, it can be seen that the statistical characteristics ($\langle n \rangle_n, Rn_n, \sqrt{\delta n_n}, RV_n$) of the CV states of certain parity generated with help of BS with $t = 0.99$ become larger in comparison with ones in figure 3. Moreover, in the overwhelming majority of cases, conditions $\langle n \rangle_n > \langle n \rangle_n$ and $\sqrt{\delta n_n} > \sqrt{\delta n_{SMSV}}$ are observed, except for cases with a small number $n$ of the subtracted photons.

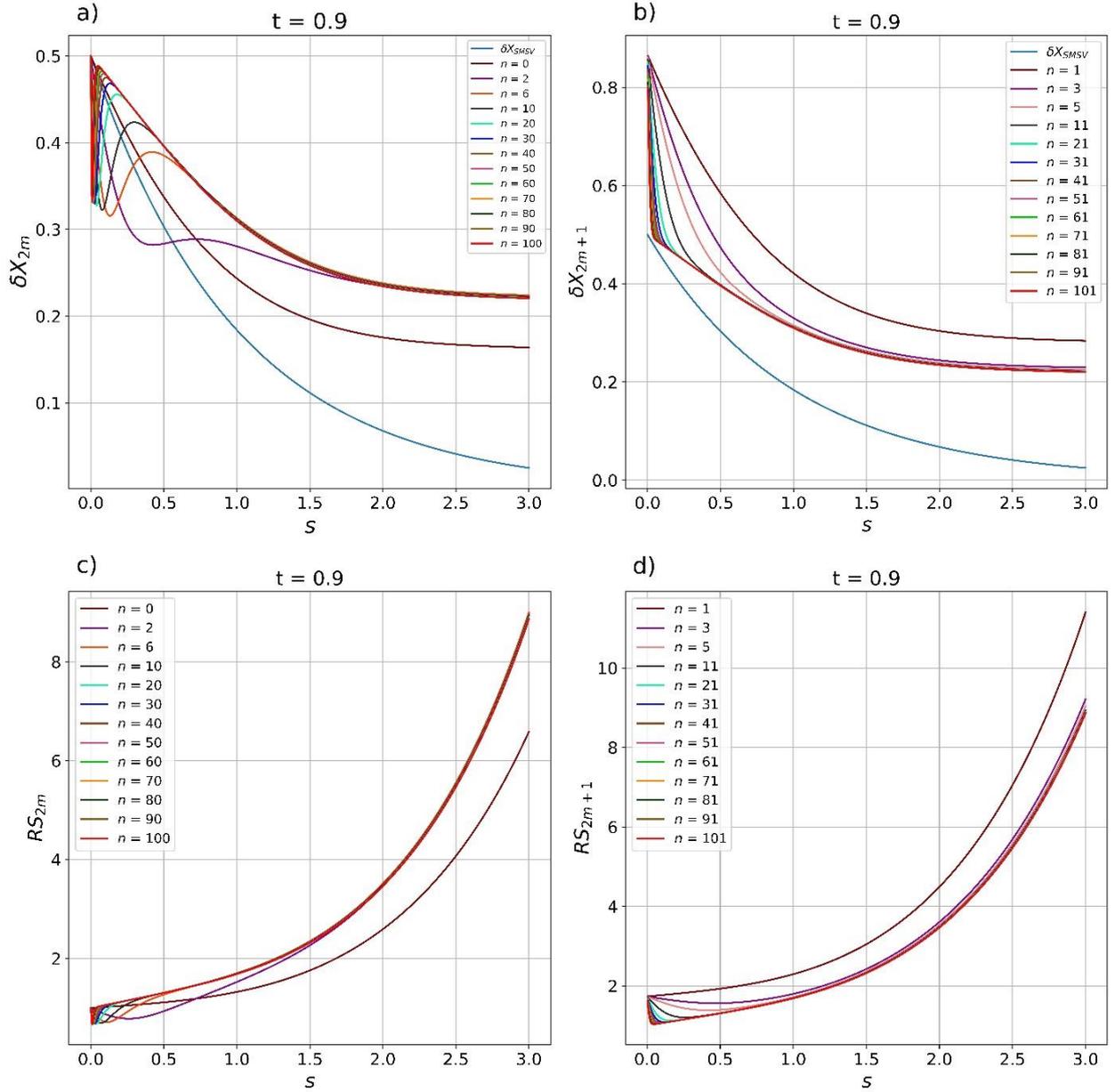

**Figure 5(a-d)** Dependencies of the quadrature component $(\delta X)_n$ for (a) $2m-$ and (b) $2m+1-$ heralded states on the squeezing amplitude $s$ for the BS parameter $t=0.9$. The quadrature component $(\delta X)_{2m}$ starts with $(\delta X)_{2m}(s=0)=0.5$, while the starting point for the $2m+1-$ heralded states is $(\delta X)_{2m+1}(s=0)=\sqrt{0.75}$. Also for comparison, the quadrature squeezing of the initial SMSV $(\delta X)_{SMSV}<0.5$ is shown (blue line). Quadrature squeezing is observed for both $2m-$ ($(\delta X)_{2m}<0.5$) and $2m+1-$ ($(\delta X)_{2m+1}<0.5$) heralded states. The ratios $RS_n=(\delta X)_n/(\delta X)_{SMSV}$ estimating how much quadrature squeezing in CV states of definite parity differs from SMSV quadrature squeezing are shown in subfigures (c) and (d).

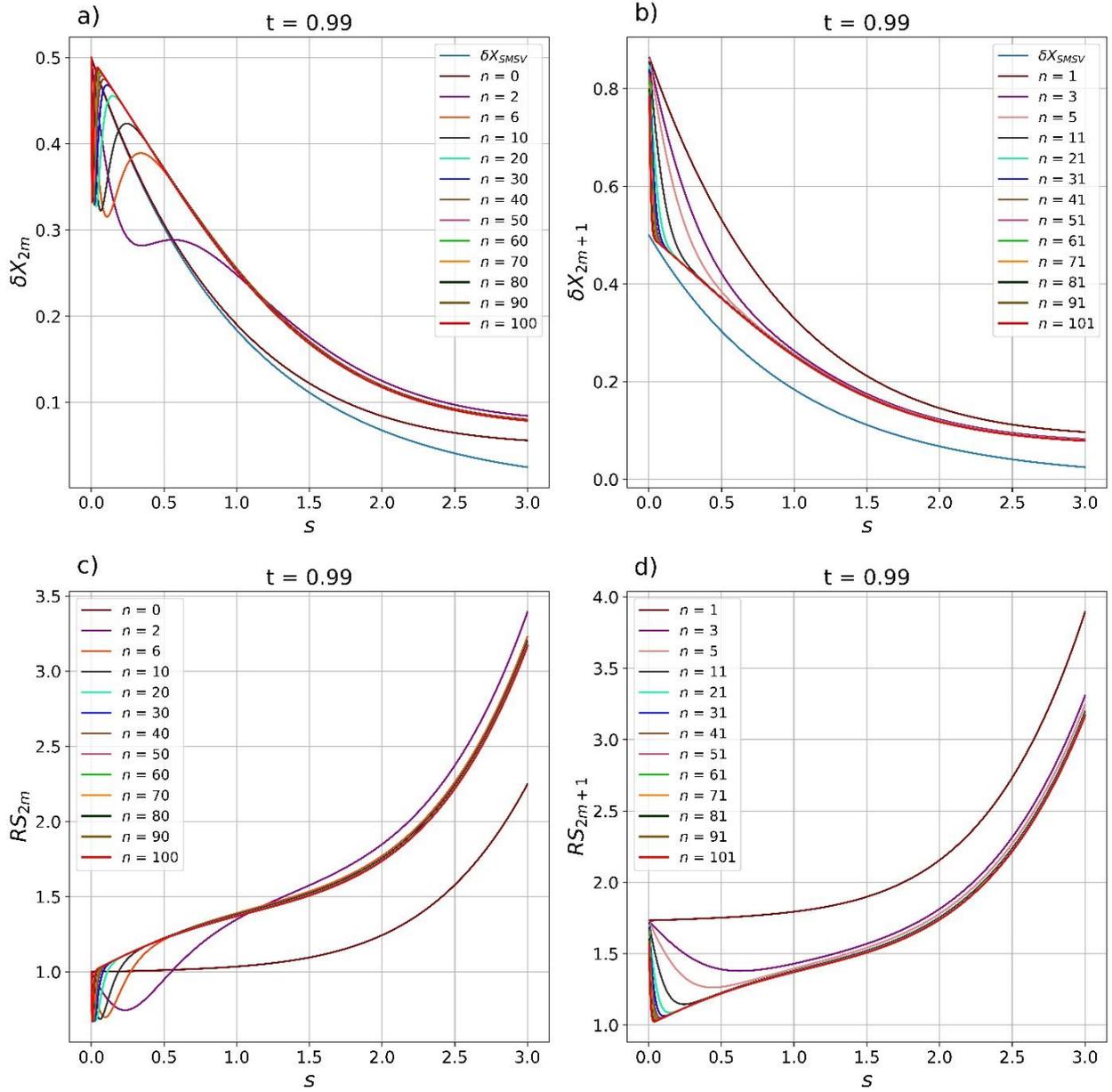

**Figure 6(a-d)** The same dependences of the quadrature components $(\delta X)_n$ and squeezing ratios $RS_n$ as in Fig. 5 on the squeezing amplitude $s$ of the original SMSV but in the case of the BS parameter $t = 0.99$. Blue lines in (a) and (c) describe SMSV quadrature squeezing.

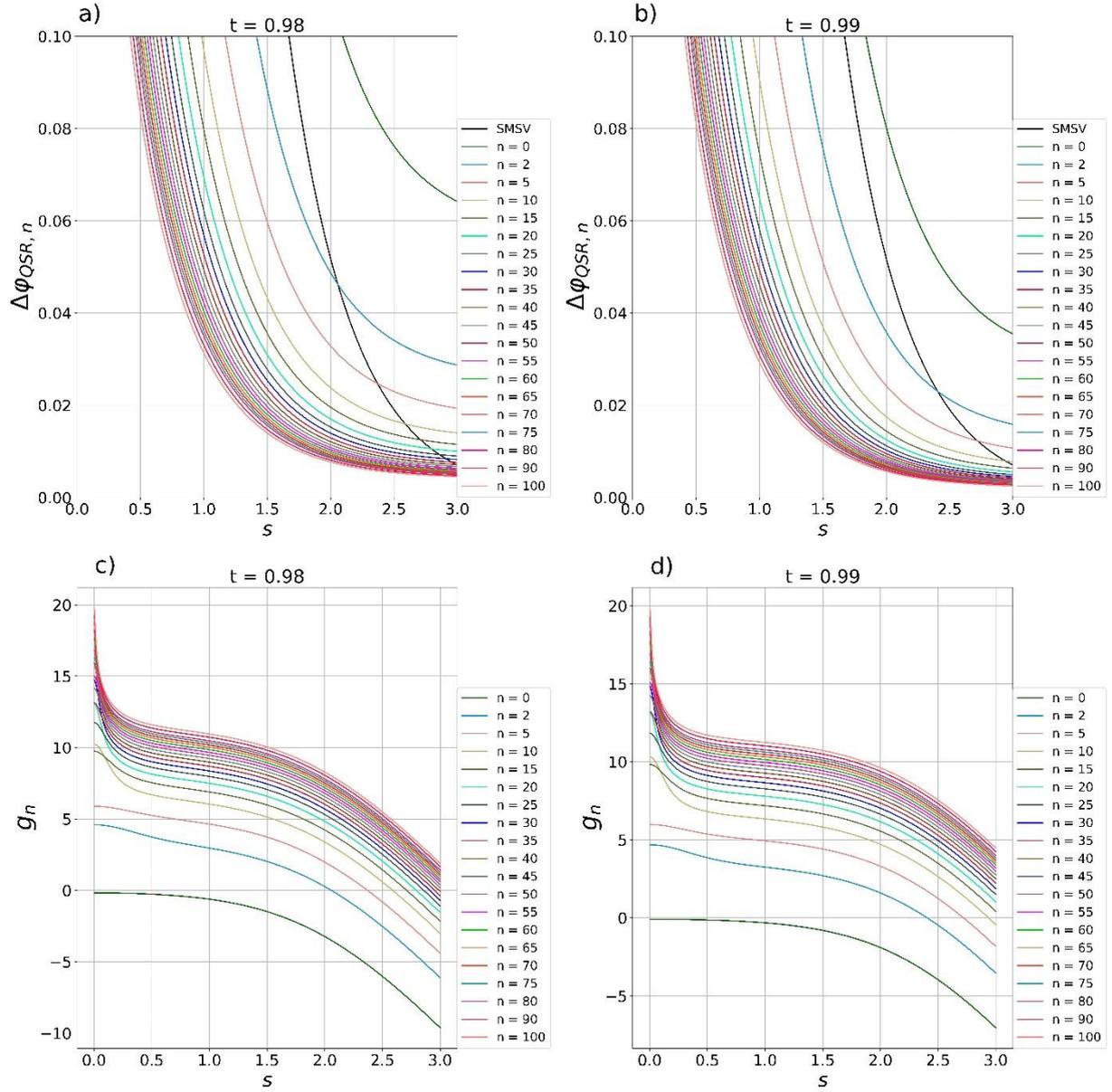

**Figure 7(a-d)** Dependence (a,b) of the QCR bound $\Delta\varphi_{QCR,n}$ of the measurement-induced states in Eq. (2,3) for observable measuring the number of photons and (c,d) sensitivity gain $g_n(dB)$ for (a,c) $t = 0.98$ and (b,d) $t = 0.99$ as a function of the squeezing amplitude $s$ of the initial SMSV state.